\documentclass[aip,preprint,superscriptaddress,longbibliography,apm]{revtex4-2}
\usepackage{graphicx, amsmath, verbatim, dsfont, amsfonts, color, amssymb, comment}
\usepackage[us]{datetime}
\usepackage{hyperref}
\usepackage{orcidlink}
\usepackage{tabularx}
\usepackage{multirow}
\usepackage{threeparttable}

\graphicspath{graphics/}

\newcommand*{\SGx}{Si$_x$Ge$_{1-x}$}
\newcommand*{\SG}{Si$_{0.15}$Ge$_{0.85}$}

\begin{document}

\title{Characterization of superconducting germanide and germanosilicide films of Pd, Pt, Rh and Ir formed by solid-phase epitaxy}

\author{Hao Li \orcidlink{0009-0004-3044-2667}}
\affiliation{Department of Physics and Astronomy, Purdue University, West Lafayette, IN 47907, USA}
\author{Zhongxia Shang \orcidlink{0000-0002-2974-2599}}
\affiliation{Purdue Electron Microscopy Center, Purdue University, West Lafayette, Indiana 47907, USA}
\affiliation{Birck Nanotechnology Center, Purdue University, West Lafayette, Indiana 47907, USA}
\author{Michael P. Lilly \orcidlink{0009-0002-5212-6230}}
\affiliation{Center for Integrated Nanotechnologies, Sandia National Laboratories, Albuquerque, NM, USA}
\author{Maksym Myronov \orcidlink{0000-0001-7757-2187}}
\affiliation{Department of Physics, University of Warwick, Coventry CV4 7AL, United Kingdom}
\author{Leonid~P.~Rokhinson \orcidlink{0000-0002-1524-0221}}
\email{leonid@purdue.edu}
\affiliation{Department of Physics and Astronomy, Purdue University, West Lafayette, IN 47907, USA}
\affiliation{Elmore Family School of Electrical and Computer Engineering, Purdue University, West Lafayette, Indiana 47907, USA}

\date{\today}

\begin{abstract}

Facilitated by recent advances in strained Ge/SiGe quantum well (QW) growth technology, superconductor–semiconductor hybrid devices based on group IV materials have been developed, potentially augmenting the functionality of quantum circuits. The formation of highly transparent superconducting platinum germanosilicide (PtSiGe) contacts to Ge/SiGe heterostructures by solid-phase epitaxy between Pt and SiGe has recently been reported, although with a relatively low critical temperature $<1\,\mathrm{K}$. Here, we present a comparative study of the superconducting properties of Pt, Pd, Rh, and Ir germanides, along with an in-depth characterization of Ir(Si)Ge films formed by solid-phase epitaxy. For films fabricated under optimal epitaxy conditions, we report $T_\mathrm{c}=3.4\,\mathrm{K}$ ($2.6\,\mathrm{K}$) for IrGe (IrSiGe). High-resolution scanning transmission electron microscopy (HRSTEM) and energy-dispersive X-ray spectroscopy (EDX) reveal that Ir reacts with Ge substrates to form a polycrystalline IrGe layer with a sharp IrGe/Ge interface.

\end{abstract}

\maketitle

\section*{Introduction}
\label{sec:intro}

The development of superconductor–semiconductor hybrid technology based on group IV semiconductors \cite{Dimoulas2006,Nishimura2007,Hendrickx2019,Vigneau2019,Aggarwal2021,Tosato2023,Valentini2024,Hinderling2024,Lakic2025} is facilitated by advances in the growth of compressively strained Ge on Si (cs-GoS), via relaxed SiGe buffer, heterostructures, where the mobility of two-dimensional hole gases in Ge quantum wells (QWs) can exceed $10^6\,\mathrm{cm^2/(V}$-s) \cite{Morrison2017,Lodari2022,Myronov2023,Myronov2023a,Wang2025}. Aluminum forms low-resistance contacts with strained Ge due to Fermi-level pinning close to the top of the valence band \cite{Dimoulas2006,Nishimura2007}, and superconductivity can be induced in Ge QWs by the proximity effect, either in direct Al/Ge contacts \cite{Hendrickx2019,Vigneau2019,Aggarwal2021}, or when Al is separated from Ge by a thin, semitransparent SiGe barrier \cite{Valentini2024}. The diffusion of Al into (Si)Ge is not self-limiting, which makes the transparency of annealed Al contacts a strong function of post-deposition processing. The sensitivity of Al contacts to thermal treatment, their low superconducting critical temperature $T_\mathrm{c}<1.5\,\mathrm{K}$, and low out-of-plane critical field $B_\mathrm{c,\perp}\sim10\,\mathrm{mT}$ have stimulated the development of alternative contacts.

To induce a hard superconducting gap in Ge, an ideal superconducting contact should have a low Schottky barrier and form a sharp, transparent interface with Ge. In CMOS technology, mono- and polycrystalline silicides are used to form low-resistance contacts between metal interconnects and Si/Ge channels \cite{Lavoie2017}, in which case, contacts are formed via a solid-phase reaction between a metal and Si/Ge. Although a number of silicides and germanides are superconducting at low temperatures \cite{Matthias1963, Raub1963, Roberts1976, Ghosh1977, Knoedler1979, Raub1984, Hirai2013, Zhang2021, Arushi2022, Nakamura2023, Strohbeen2024}, many compounds with high $T_\mathrm{c}>4$ K, such as NbGe$_2$ \cite{Ghosh1977,Knoedler1979,Zhang2021}, form at very high temperatures beyond the $\sim 500\ ^\circ$C thermal budget of strained Ge/SiGe heterostructures \cite{Gaudet2006}. Some platinum group metals form silicides and germanides at low temperatures ($<500\ ^\circ$C), and the formation of superconducting PtGe ($T_\mathrm{c}=0.4$ K) \cite{Matthias1963,Raub1963,Roberts1976,Raub1984} and RhGe ($T_\mathrm{c}=0.96$-1.7 K) \cite{Matthias1963,Raub1963,Roberts1976,Raub1984,Hirai2013} has been demonstrated. Notably, polycrystalline platinum germanosilicide (PtSiGe) contacts have been shown to induce a hard superconducting gap in shallow Ge QWs with high transparency $\tau\sim0.95$ \cite{Tosato2023,Hinderling2024}. A high out-of-plane upper critical magnetic field $B_\mathrm{c2,\perp}=0.9$ T has been measured in hybrid devices with small PtSiGe leads \cite{Lakic2025}, substantially higher than $B_\mathrm{c2,\perp}\sim0.1$ T in large contacts \cite{Tosato2023,Lakic2025}. Another transition metal, Ta, has recently been found to form a solid solution of Ta$_5$Ge$_3$ and TaGe$_2$ when grown via molecular beam epitaxy (MBE) on Ge substrates at $T=400\ ^\circ$C \cite{Strohbeen2024}. These films exhibit $T_\mathrm{c}\sim 1.8$-2 K, $B_\mathrm{c2,\perp}\sim 1.88$ T, and $B_\mathrm{c2,||}\sim 5.1$ T, in agreement with previous studies of tantalum germanide \cite{Ghosh1977,Li2024}. 

Within the platinum group germanides, Ir-based compounds show the highest $T_\mathrm{c}$: IrGe ($T_\mathrm{c}=4.7$–$5.3\,\mathrm{K}$, $B_\mathrm{c1}(0)=13.3\,\mathrm{mT}$, $B_\mathrm{c2}(0)=0.82$–$1.13\,\mathrm{T}$) \cite{Matthias1963,Raub1963,Roberts1976, Raub1984,Hirai2013,Arushi2022}, Ir$_3$Ge$_7$ ($T_\mathrm{c}=0.87\,\mathrm{K}$) \cite{Raub1963,Roberts1976}, and IrGe$_4$ ($T_\mathrm{c}=1.22\,\mathrm{K}$, $B_\mathrm{c2}=11.5$–$22.5\,\mathrm{mT}$) \cite{Nakamura2023}. Based on results from specific heat and muon spin relaxation experiments, an enhanced $T_\mathrm{c}$ of IrGe is attributed to the presence of low-lying phonons \cite{Hirai2013, Arushi2022}. Although single IrGe crystals are grown at $T=700\,^\circ\mathrm{C}$ \cite{Hirai2013}, there is some evidence that the onset of solid-phase reaction between Ir and Ge begins at $400\,^\circ\mathrm{C}$ \cite{Habanyama2018}. While IrGe shows an enhanced $T_\mathrm{c}$, IrSi is not known to be a superconductor down to mK temperatures. The superconducting properties of IrSiGe have not been investigated.

In this paper, we compare the superconducting properties of thin films formed by solid-phase epitaxy of Pd, Pt, Rh, and Ir on Ge (001) or relaxed \SG\ layers grown on Si(001), and present a systematic study of Ir-based films. We show that within the temperature range of $400$–$500\,^\circ\mathrm{C}$ (within the thermal budget of strained Ge/SiGe heterostructures), self-limiting polycrystalline films of IrGe and IrSiGe are formed by solid-phase epitaxy. For optimally grown IrGe (IrSiGe) films, we report $T_\mathrm{c}\sim3.4\,\mathrm{K}$ ($2.6\,\mathrm{K}$) and $B_\mathrm{c2,\perp}\sim3.4\,\mathrm{T}$ ($2.8\,\mathrm{T}$), the latter being significantly larger than $B_\mathrm{c}=0.82$–$1.13\,\mathrm{T}$ reported for bulk IrGe \cite{Hirai2013,Arushi2022}. High-resolution scanning transmission electron microscopy (HRSTEM) of optimally annealed Ir/Ge films reveals the formation of IrGe microcrystals with sharp boundaries between IrGe and Ge.

\section*{Results}
\label{sec:results}

In this study, we use three types of substrates: undoped Ge (001) single-crystal wafers (\textit{i}-Ge); strain-relaxed \SG\ thick films capped with $2\,\mathrm{nm}$ of Ge, grown epitaxially on strain-relief buffers on Si(001) wafers (SiGe); and B-doped ($>10^{20}\,\mathrm{cm^{-3}}$) Ge layers grown on relaxed Ge (\textit{p}-Ge). All heterostructures are grown by reduced pressure chemical vapor deposition (RP-CVD) \cite{Myronov2023,Myronov2023a}. Prior to metal deposition, the wafers are cleaned using acetone, isopropanol, and deionized water, followed by a dip in diluted HF and HCl to remove native oxides. We expect that most of the $2\,\mathrm{nm}$ Ge capping layer on SiGe is naturally oxidized and is removed during the last step of acid treatment. Metal films are deposited either by DC sputtering (Ir) or electron-beam evaporation (Pt, Pd, Rh) onto substrates held at room temperature in an ultra-high vacuum deposition systems with a base pressure of $\sim10^{-9}\,\mathrm{Torr}$. After metal deposition, the samples are annealed in a custom-built rapid thermal annealer (RTA) with precise temperature control, in an Ar atmosphere. Transport measurements are carried out in either a dilution refrigerator or a $^3$He cryostat using standard four-probe low-frequency lock-in techniques.

\begin{figure}[htbp]
    \centering
    \includegraphics[width=0.48\textwidth]{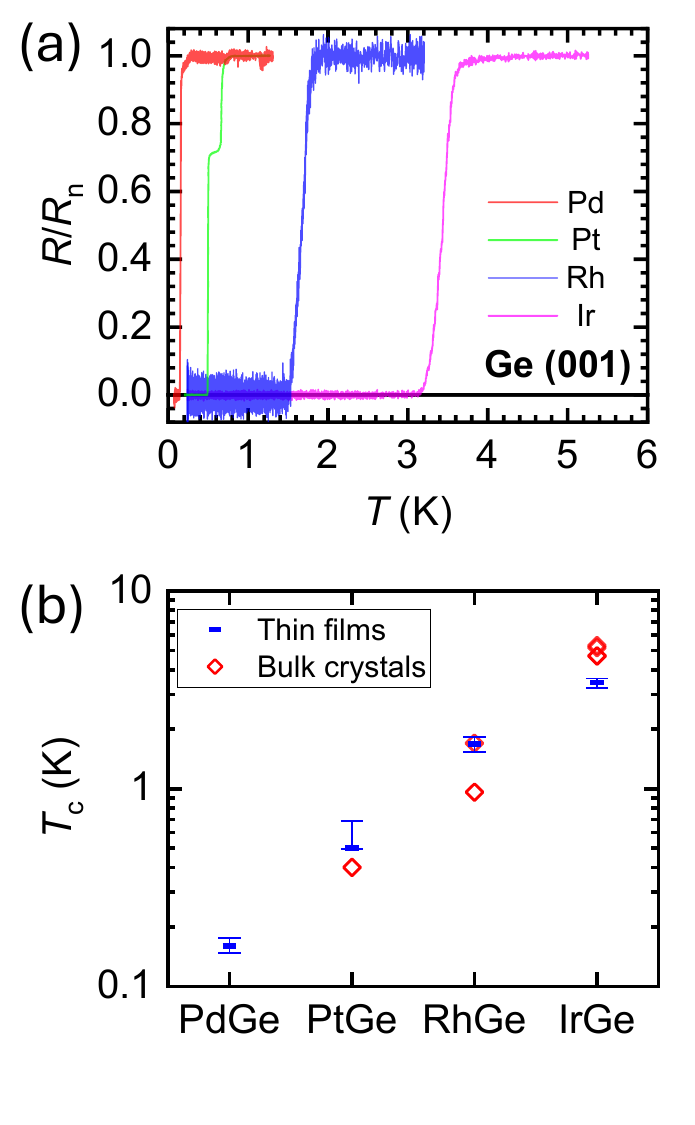}
    \caption{(a) Temperature dependence of the normalized resistance $R/R_\mathrm{n}$ for optimally annealed $10\,\mathrm{nm}$ Pd, Pt, Rh, and Ir films deposited on Ge (001) substrates. (b) $T_\mathrm{c}$ for thin films in this study (blue) are compared with $T_\mathrm{c}$ for bulk crystals (red) from Refs.~\cite{Matthias1963,Raub1963,Roberts1976, Raub1984,Hirai2013,Arushi2022}. Vertical bars indicate the width of the transition region defined as $0.05<R/R_\mathrm{n}<0.95$.}
    \label{fig:PdPtRhIrGeRT}
\end{figure}

In \hyperref[fig:PdPtRhIrGeRT]{Fig.~1(a)}, the temperature dependence of the normalized resistance is plotted for Pd, Pt, Rh, and Ir germanides formed under optimal thermal treatment conditions of $10\,\mathrm{nm}$ films. The superconducting transition temperatures $T_\mathrm{c}$, defined here as the midpoint of the transition where the resistance drops by a factor of two relative to the normal resistance ($R=0.5 R_\mathrm{n}$), follow the trend of bulk germanides. For the growth of Pd and Pt germanides, we used \textit{p}-Ge substrates, while Rh and Ir germanides were grown on \textit{i}-Ge. 

Previous studies of diffusion kinetics using X-ray diffraction (XRD) and TEM techniques indicate that the solid-phase reaction between Pd and Ge initially proceeds via the diffusion-controlled growth of Pd$_2$Ge, which is converted into PdGe at $T>150\,^\circ\mathrm{C}$; PdGe phase remains stable up to its melting point $T\approx740\,^\circ\mathrm{C}$ \cite{Wittmer1977,Ottaviani1977,Majni1977,Majni1978,Ottaviani1979,Hsieh1988,Chen2004,Geenen2014,PerrinToinin2016,Habanyama2018,Habanyama2018a,Krause2023}. Thus, we expect that the PdGe phase dominates in our samples for annealing at $T\geqslant300\,^\circ\mathrm{C}$, and the measured $T_\mathrm{c}=0.16\,\mathrm{K}$ reflects the superconducting properties of PdGe films. For annealed Pt/Ge films the measured $T_\mathrm{c}=0.5\,\mathrm{K}$ is higher than the reported $T_\mathrm{c}=0.4\,\mathrm{K}$ for bulk PtGe \cite{Matthias1963}. Earlier studies using XRD indicate the formation of the PtGe phase when films are annealed in the $380$–$440\,^\circ\mathrm{C}$ temperature range \cite{Gaudet2006}. For Rh/Ge, we measure $T_\mathrm{c}=1.68\,\mathrm{K}$, close to the bulk RhGe $T_\mathrm{c}\sim1.7\,\mathrm{K}$ \cite{Hirai2013}. Rh$_5$Ge$_3$ orthorhombic phase has higher $T_\mathrm{c}=2.12\,\mathrm{K}$ \cite{Matthias1963,Raub1963,Roberts1976,Raub1984}, but this phase has not been found in previous studies of the solid-phase reaction of Rh/Ge thin films \cite{Wittmer1977,Gaudet2006,Habanyama2018}. RhGe$_4$ phase also has higher $T_\mathrm{c}=2.55\,\mathrm{K}$, but this phase is only stabilized under high pressure \cite{Nakamura2023} and we do not expect it to be formed in our experiments. For Ir/Ge, the measured $T_\mathrm{c}=3.44\,\mathrm{K}$ is slightly less than the $4.7$–$5.3\,\mathrm{K}$ reported for stoichiometric IrGe crystals \cite{Matthias1963,Hirai2013}, but higher than that for Ir$_3$Ge$_7$ ($T_\mathrm{c}=0.87\,\mathrm{K}$ \cite{Raub1963}) or IrGe$_4$ ($T_\mathrm{c}=1.22\,\mathrm{K}$ \cite{Nakamura2023}) phases. 

The superconducting properties of Pd, Pt, and Ir germanosilicides are found to reflect the composition of the \SGx\ substrate. For films grown on \SG\ substrates, we measure $T_\mathrm{c}=0.30\,\mathrm{K}$ for Pd/SiGe, $T_\mathrm{c}=0.59\,\mathrm{K}$ for Pt/SiGe, and $T_\mathrm{c}=2.56\,\mathrm{K}$ for Ir/SiGe. These values are close to $T_\mathrm{c}^\mathrm{MSiGe}\approx 0.15\times T_\mathrm{c}^\mathrm{MSi}+0.85\times T_\mathrm{c}^\mathrm{MGe}$ (M = Pd, Pt, or Ir) [Fig.~S1 in the \hyperref[supp]{supplementary material}]. The observation of a weighted $T_\mathrm{c}$ suggests the formation of a proportional mixture of germanides and silicides. 

\begin{figure*}[t]
\includegraphics[width=\textwidth]{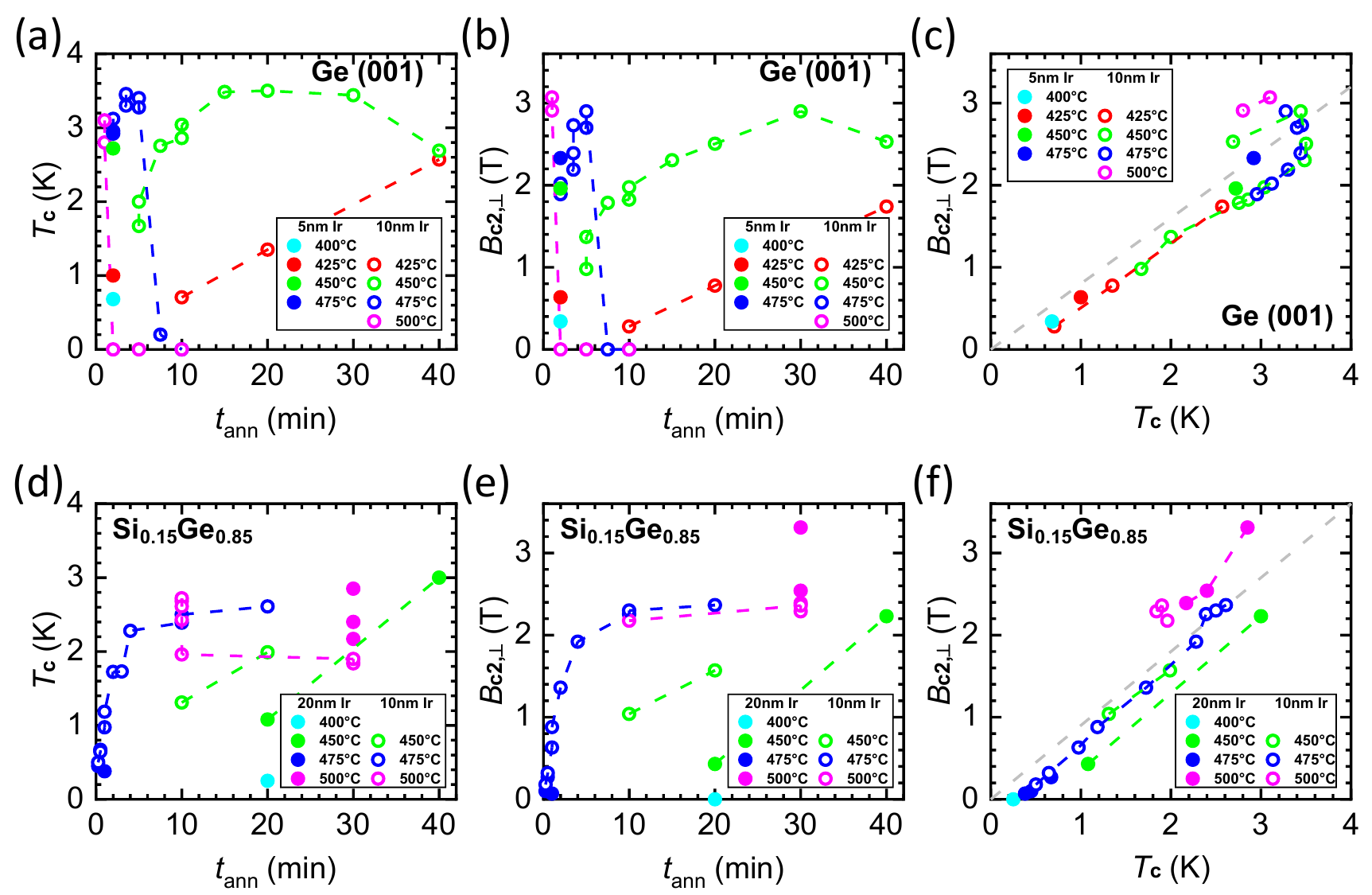}
\caption{(a,b,d,e) The critical temperature $T_\mathrm{c}$ and the out-of-plane critical field $B_\mathrm{c2,\perp}$ are plotted as a function of annealing time $t_\mathrm{ann}$ for \textit{i}-Ge and SiGe substrates, showing their dependence on the Ir film thickness and annealing temperature. $T_\mathrm{c}<0.25\,\mathrm{K}$ is plotted as $T_\mathrm{c}=0$. The filled (open) circles correspond to different Ir thicknesses indicated in the legends, and colors differentiate different annealing temperatures. In (c,f) $B_\mathrm{c2,\perp}$ is re-plotted as a function of $T_\mathrm{c}$.}
\label{fig:IrGeIrSiGeTcBc}
\end{figure*} 

The dependence of $T_\mathrm{c}$ and $B_\mathrm{c2}$ (defined at a midpoint of resistance drop, $R(B_\mathrm{c2})=0.5 R_\mathrm{n}$) of Ir/Ge and Ir/SiGe films on Ir thickness, annealing time $t_\mathrm{ann}$, and annealing temperature is summarized in \hyperref[fig:IrGeIrSiGeTcBc]{Fig.~2}. The initial increase of $T_\mathrm{c}$ and $B_{\mathrm{c2},\perp}$ with $t_\mathrm{ann}$ is attributed to the finite barrier for Ir diffusion into \textit{i}-Ge and SiGe substrates. The corresponding activation energy can be found by analyzing the time $t_\mathrm{ann}^0$ needed to anneal a film to reach the same $T_\mathrm{c}$ for different annealing temperatures $T_\mathrm{ann}$. The $t_\mathrm{ann}^0(T_\mathrm{ann})$ dependence for $T_\mathrm{c}\sim(2.8\pm0.2)\,\mathrm{K}$ is found to be exponential, and from the Arrhenius plot we extract the activation energy $E_\mathrm{a}\approx2.7\,\mathrm{eV}$ assuming $t_\mathrm{ann}^0\propto1/D$ and a diffusion coefficient $D\propto \exp(-E_\mathrm{a}/k_\mathrm{B}T_\mathrm{ann})$, where $k_\mathrm{B}$ is the Boltzmann constant [Fig.~S2 in the \hyperref[supp]{supplementary material}]. Here we assume that partially formed germanides are morphologically similar and neglect the reduction of $T_\mathrm{c}$ due to the reverse proximity effect in the presence of metallic Ir in under-annealed samples. Samples with thinner Ir films require shorter annealing time to reach the same $T_\mathrm{c}$.

\begin{figure*}[t]
\includegraphics[width=\textwidth]{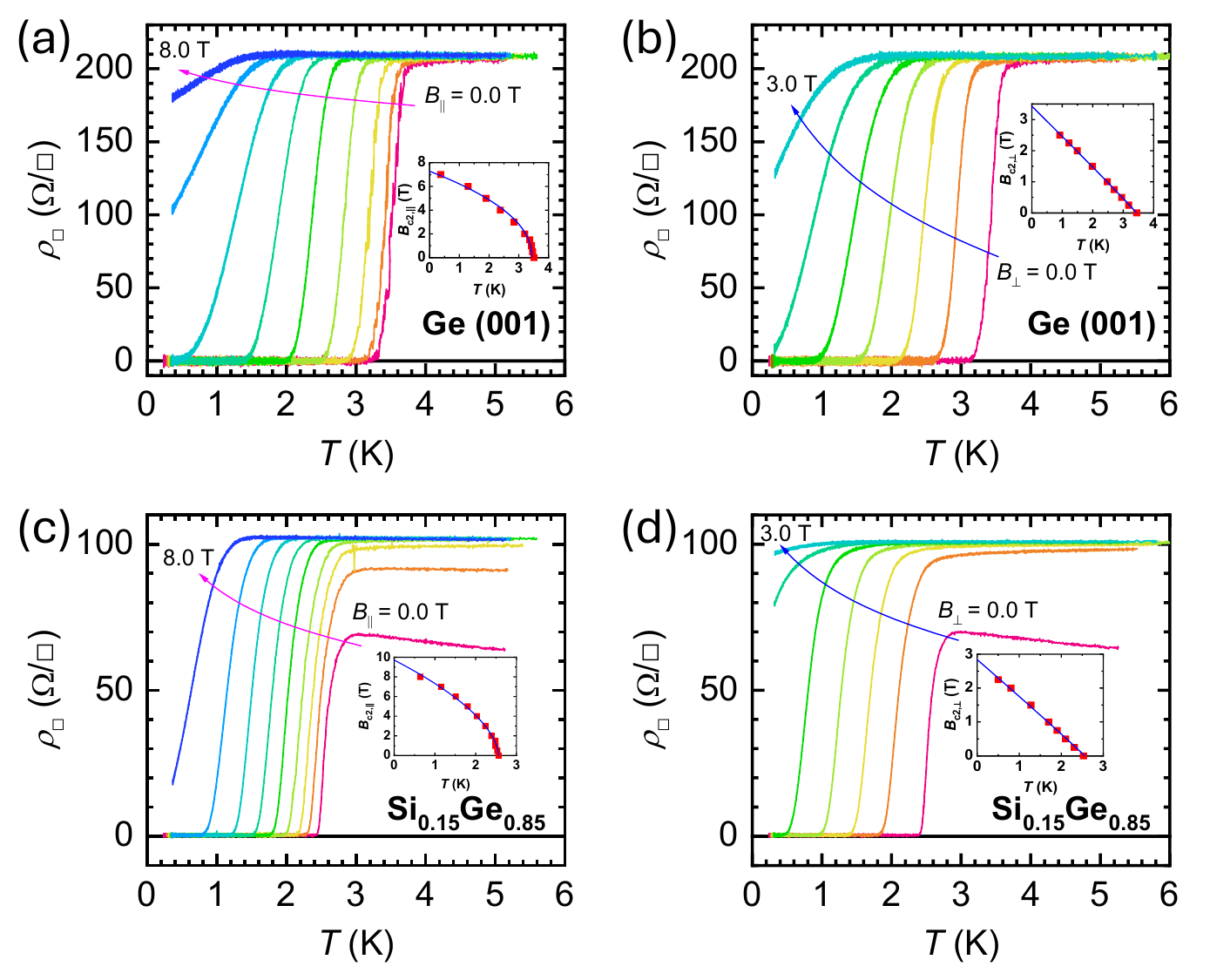}
\caption{The change in resistivity as a function of temperature is plotted for various in-plane $B_{||}$ and out-of-plane $B_\perp$ magnetic fields for the optimally annealed Ir/\textit{i}-Ge and Ir/SiGe films. The insets show the temperature dependence of $B_{\mathrm{c2}}$, and the blue lines represent fits using \hyperref[eq:GLBcTc]{Eq.~1}.}
\label{fig:IrGeIrSiGeRTB}
\end{figure*} 

\begin{table}[t]
\begin{center}
\caption{$T_\mathrm{c}$ is an experimental value for the films studied in \hyperref[fig:IrGeIrSiGeRTB]{Fig.~3}, while $B_\mathrm{c2}$ and $\alpha$ are extracted from the fits to \hyperref[eq:GLBcTc]{Eq.~1}. $B_\mathrm{P}$ is an estimated Pauli limit as discussed in the text. Superconducting coherence length $\xi(0)$ and effective film thickness $d_\mathrm{eff}$ are estimated from the in-plane (IP) and out-of-plane (OOP) critical fields $B_\mathrm{c2}$. Estimation of the Ginzburg–Landau coherence length $\xi_0$ and mean free path $l$ are discussed in the text.}
\label{tab:IrGeIrSiGeGLfit}
\begin{tabularx}{0.45\textwidth}{c>{\centering\arraybackslash}X>{\centering\arraybackslash}X@{\hspace{0.01\textwidth}}>{\centering\arraybackslash}X>{\centering\arraybackslash}X}
\hline\hline
 & \multicolumn{2}{c}{Ir/Ge} & \multicolumn{2}{c}{Ir/SiGe}\\
 & IP & OOP & IP & OOP\\
\hline
$T_\mathrm{c}$ (K) & \multicolumn{2}{c}{3.44} & \multicolumn{2}{c}{2.56}\\
$B_\mathrm{c2}(0)$ (T) & 7.28 & 3.43 & 9.71 & 2.84\\
$\alpha$ & 0.46 & 0.97 & 0.57 & 0.98\\
$B_\mathrm{P}$ (T) & \multicolumn{2}{c}{8.3} & \multicolumn{2}{c}{6.2}\\
$\xi(0)$ (nm) & \multicolumn{2}{c}{9.8} & \multicolumn{2}{c}{10.8}\\
$d_\mathrm{eff}$ (nm) & \multicolumn{2}{c}{16.0} & \multicolumn{2}{c}{10.9}\\
$l$ (nm) & \multicolumn{2}{c}{1.7} & \multicolumn{2}{c}{\textbf{--}}\\
$\xi_0$ (nm) & \multicolumn{2}{c}{56.5} & \multicolumn{2}{c}{\textbf{--}}\\
\hline\hline
\end{tabularx}
\end{center}
\end{table} 

For Ir on \textit{i}-Ge samples, after an initial increase with annealing time, $T_\mathrm{c}$ reaches a maximum and then decreases with further annealing. A sharp decrease in $T_\mathrm{c}$ is observed for samples annealed at $T\geqslant475\,^\circ\mathrm{C}$, where an extra minute of annealing results in a drop in $T_\mathrm{c}$ from $3.2\,\mathrm{K}$ to $<0.25\,\mathrm{K}$. For IrGe films, the optimal annealing temperature is found to be $450\,^\circ\mathrm{C}$; this $T_\mathrm{ann}$ provides a large $t_\mathrm{ann}\approx 12$–$30\,\mathrm{min}$ window to form films with the highest $T_\mathrm{c}$. As we will discuss below, optimally annealed films consist of IrGe microcrystals.

For Ir on SiGe, no significant decrease in $T_\mathrm{c}$ for long annealing times has been observed. We speculate that the reduction in $T_\mathrm{c}$ in over-annealed Ir/Ge samples is due to the formation of a competing semiconducting Ir$_4$Ge$_5$ phase with lower formation energy \cite{Jain2013,Saal2013,Kirklin2015}. This explanation is consistent with the observed higher stability of IrSiGe films, since IrSi is expected to have a lower formation energy than the competing Ir$_3$Si$_4$ and Ir$_3$Si$_5$ metallic phases. The formation of a semi-insulating phase in over-annealed Ir/Ge samples is also consistent with a sharp increase in resistivity and a semiconducting $T$-dependence (slight resistance increase below $1\,\mathrm{K}$) [Table.~S2 and Fig.~S3 in the \hyperref[supp]{supplementary material}].

The critical field $B_\mathrm{c2,\perp}$, plotted in \hyperref[fig:IrGeIrSiGeTcBc]{Figs.~2(b)} and \hyperref[fig:IrGeIrSiGeTcBc]{2(e)}, evolves similarly to $T_\mathrm{c}$ as a function of Ir thickness, annealing time, and annealing temperature. This similarity is emphasized by an almost linear dependence $B_\mathrm{c2,\perp}\propto T_\mathrm{c}$, shown in \hyperref[fig:IrGeIrSiGeTcBc]{Figs.~2(c)} and \hyperref[fig:IrGeIrSiGeTcBc]{2(f)}, where $B_\mathrm{c2,\perp}$ is plotted as a function of $T_\mathrm{c}$ for all devices and annealing conditions. The highest $B_\mathrm{c2,\perp}=3.07\,\mathrm{T}$ measured in Ir/Ge and $B_\mathrm{c2,\perp}=3.31\,\mathrm{T}$ measured in Ir/SiGe films are significantly higher than the $B_\mathrm{c2,\perp}=0.82$–$1.13\,\mathrm{T}$ reported for bulk IrGe crystals \cite{Hirai2013,Arushi2022}.

In \hyperref[fig:IrGeIrSiGeRTB]{Fig.~3}, the temperature dependence of resistivity is plotted for different in-plane ($B_{||}$) and out-of-plane ($B_\perp$) magnetic fields. In the insets, the temperature dependencies of $B_\mathrm{c2,\perp}$ and $B_\mathrm{c2,||}$ are fitted to
% the Ginzburg-Landau (GL) relation: 
\begin{equation}
    B_\mathrm{c2}(T)=B_\mathrm{c2}(0)\left(1-\frac{T}{T_\mathrm{c}}\right)^\alpha,
    \label{eq:GLBcTc}
\end{equation}
where $\alpha$ and $B_\mathrm{c2}(0)$ are the fitting parameters summarized in \hyperref[tab:IrGeIrSiGeGLfit]{Table.~I}. For the out-of-plane field $\alpha\approx 1$, while for the in-plane field $\alpha\approx 0.5$, confirming the two-dimensional nature of IrGe and IrSiGe films with $d_\mathrm{eff}\lesssim 1.8\xi$. The temperature dependence of the critical fields, $B_\mathrm{c2,\bot}(T)=\Phi_0/(2\pi\xi^2)$ and $B_\mathrm{c2,||}(T)\approx\sqrt{3}\Phi_0/(\pi\xi d_\mathrm{eff})$, originates from the temperature dependence of the coherence length, $\xi(T)\approx\xi(0)/\sqrt{1-T/T_\mathrm{c}}$, where in a dirty superconductor $\xi(0)=\sqrt{\xi_0 l}$ is a geometrical average of a Ginzburg–Landau coherence length $\xi_0$ and a mean free path $l$, $\Phi_0=h/2e$ is the magnetic flux quantum \cite{Tinkham1996}. The mean free path, $l = \hbar \sqrt[3]{3\pi^2}n^{2/3} / (e^2 \rho)$, is calculated using the resistivity $\rho$ and electron density $n$ measured in the normal state just above $T_\mathrm{c}$. The ratio $\xi_0/l = 33.2$ indicates dirty-limit superconductivity in IrGe. We note that the critical in-plane field is comparable to (IrGe), or exceeds (IrSiGe), the Pauli limit $B_\mathrm{P}=\Delta_0/(\sqrt{2}\mu_\mathrm{B})$, where $\Delta_0\approx 2.3 k_\mathrm{B} T_\mathrm{c}$ is the zero-temperature superconducting gap for Ir(Si)Ge, and $\mu_\mathrm{B}$ is the Bohr magneton, indicating strong spin–orbit interactions \cite{Bruno1973}. The high $B_\mathrm{c2,\perp}$ values, $B_{c2,\perp}(\mathrm{IrGe})=3.43\,\mathrm{T}$ and $B_\mathrm{c2,\perp}(\mathrm{IrSiGe})=2.84\,\mathrm{T}$, significantly exceed the $1.13\,\mathrm{T}$ reported in bulk IrGe \cite{Arushi2022}. This enhancement can be attributed to the microcrystalline nature of the films. These high $B_\mathrm{c2,\perp}$ values could be useful for studying the interplay between superconductivity and quantum Hall states in high-mobility Ge two-dimensional hole gases. 

\begin{figure*}[t]
\includegraphics[width=\textwidth]{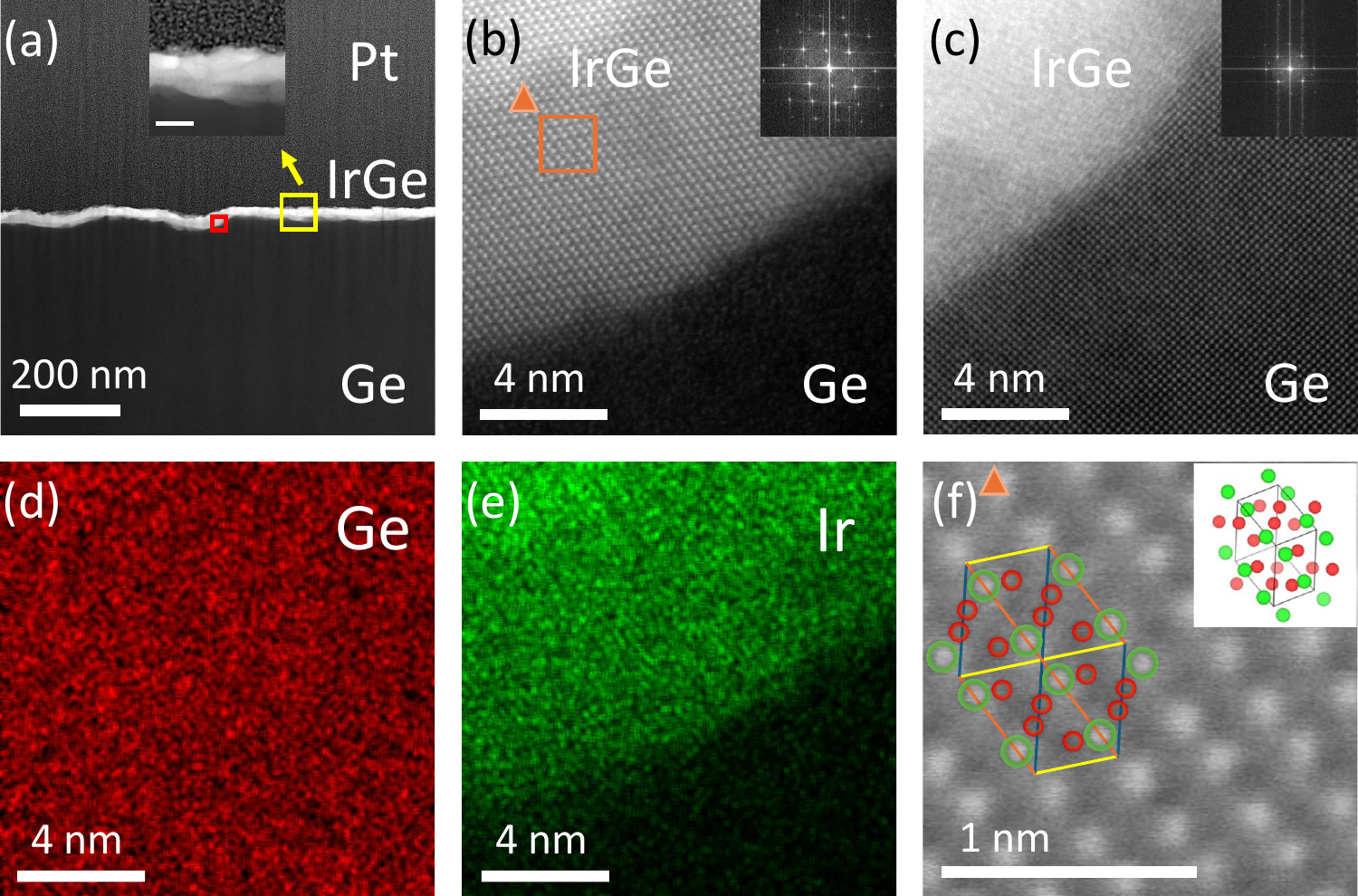}
\caption{(a) Low magnification cross-sectional STEM image of an Ir $10\,\mathrm{nm}$/Ge (001) sample annealed at $475\,^\circ\mathrm{C}$ for $3.5\,\mathrm{min}$. The sample is coated with Pt protection layer. Inset: magnified image shows microcrystalline composition of the IrGe film, the scale bar is $20\,\mathrm{nm}$. (b,c) High-angle annular dark field scanning transmission electron microscopy (HAADF-STEM) near IrGe/Ge interface is performed at slightly different angles to match either [111] orientation of IrGe or [100] orientation of Ge lattices. Insets: fast Fourier transforms (FFT) of the images. (d,e) Energy-dispersive X-ray spectroscopy (EDX) of (b). (f) Atomic resolution STEM image of IrGe (orange square in (b)) with red and green circled indicating positions of Ir and Ge atoms correspondingly. Yellow, blue and orange lines indicate $a$, $b$ and $c$ axes of the IrGe lattice. Inset: a view of an IrGe lattice seen along [111] direction. }
\label{fig:IrGeSTEM111}
\end{figure*}

High-resolution scanning transmission electron microscopy analysis of a $10\,\mathrm{nm}$ Ir/\textit{i}-Ge film annealed at $475\,^\circ\mathrm{C}$ for $3.5\,\mathrm{min}$ ($T_\mathrm{c}\sim3.4\mathrm{K}$) is presented in \hyperref[fig:IrGeSTEM111]{Fig.~4}. Low-resolution images show the formation of a uniform film with a thickness of $d_\mathrm{eff}=(24\pm 3)\,\mathrm{nm}$. The density of IrGe, $13.10\,\mathrm{g/cm^3}$, can be calculated from its structural parameters \cite{Hirai2013, Arushi2022}. Assuming that all Ir reacts with Ge to form the IrGe phase, $10\,\mathrm{nm}$ of Ir should react with $16\,\mathrm{nm}$ of Ge to form $23.7\,\mathrm{nm}$ of IrGe. This value is in excellent agreement with the film thickness obtained from wide-area STEM images, suggesting that most Ir reacts with Ge to form IrGe. 

This conclusion is corroborated by cross-sectional aberration-corrected high-angle annular dark-field scanning transmission electron microscopy (HAADF-STEM) images, which show the formation of IrGe microcrystals with sizes of 10–$20\,\mathrm{nm}$ throughout the thickness of the film. In \hyperref[fig:IrGeSTEM111]{Figs.~4(b)} and \hyperref[fig:IrGeSTEM111]{4(c)}, an IrGe/Ge boundary is shown at two slightly different angles, highlighting the crystalline structure of an IrGe crystal and the surrounding Ge matrix. The boundary is sharp, with a negligibly small concentration of Ir in Ge, as evidenced by energy-dispersive X-ray spectroscopy (EDX) analysis of Ir and Ge distribution across the boundary \hyperref[fig:IrGeSTEM111]{Figs.~4(d)} and \hyperref[fig:IrGeSTEM111]{4(e)}. Note that the STEM specimen is $\lesssim100\,\mathrm{nm}$ thick, a few times larger than the average IrGe grain size. Unlike STEM, which probes only a few nanometers from the surface, EDX averages the Ir concentration over the entire thickness of the film. Near an IrGe/Ge interface, EDX may pick up a signal from an overlapping IrGe microcrystal at the back of the specimen not visible in the STEM image, which can explain some small Ir background detected in the otherwise stoichiometrically perfect Ge matrix. 

Detailed analysis of an atomic-resolution image of IrGe microcrystals reveals the formation of an orthorhombic IrGe phase (Pnma, space group No. 62) with lattice parameters $b = (5.607 \pm 0.028)\,\text{\r{A}}$, $c = (6.294 \pm 0.018)\,\text{\r{A}}$ and angle $(89.62\pm 0.34)^\circ$ between $b$ and $c$. Lattice parameters determined for bulk IrGe crystals using XRD \cite{Hirai2013, Arushi2022} fall within the error bars of our measurements, indicating the formation of unstrained IrGe.

\begin{figure}[htbp]
\includegraphics[width=0.48\textwidth]{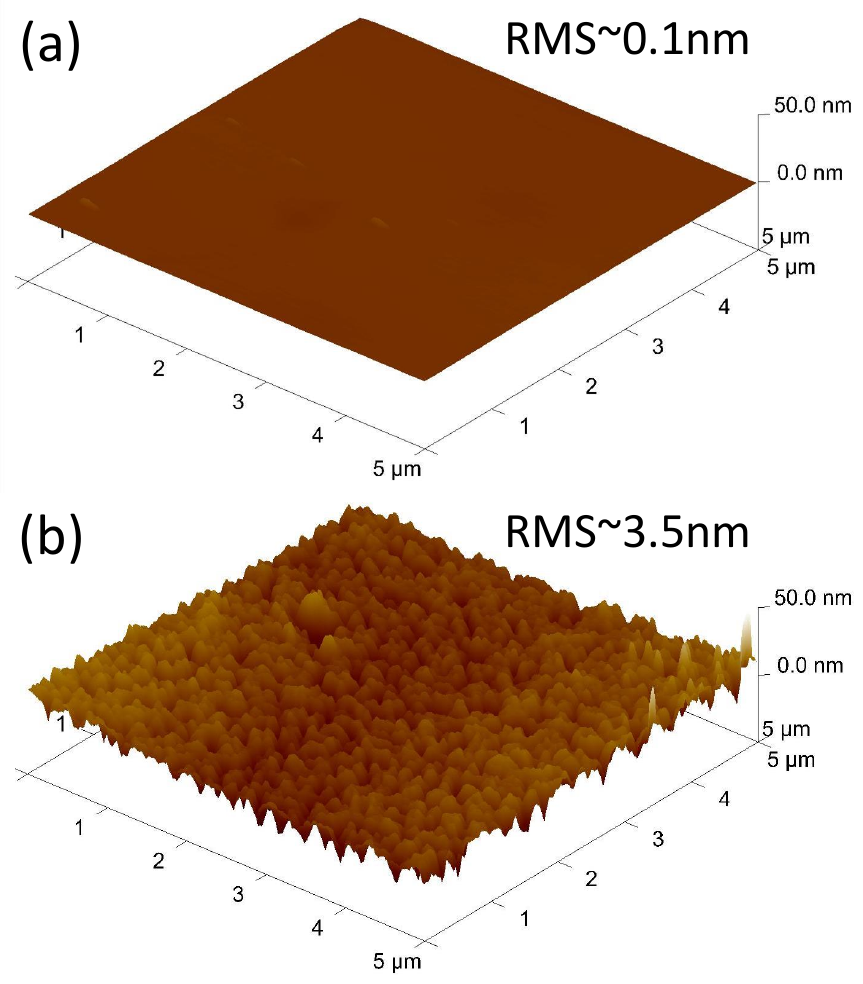}
\caption{AFM images of 10 nm Ir on \textit{i}-Ge before (a) and after (b) annealing at $475\,^\circ\mathrm{C}$ for $3.5\,\mathrm{min}$.}
\label{fig:IrGeAFM}
\end{figure}

Surface morphology studies of IrGe films corroborate the formation of a polycrystalline film. In \hyperref[fig:IrGeAFM]{Fig.~5}, we plot atomic force microscope (AFM) images of a sample before and after annealing. The surface of an Ir/Ge (001) sample before annealing is smooth with an RMS roughness $R_\mathrm{q}\sim0.1\,\mathrm{nm}$. Some shallow holes (a few nm deep) likely originate from the surface of the Ge wafer and formed during acid cleaning prior to Ir deposition. Surface of the annealed sample shows formation of bumps with a lateral size $(57\pm 22)$ nm and $R_\mathrm{q}\sim(3.5\pm1.0)\,\mathrm{nm}$. Based on TEM images, these bumps can be attributed to the formation of IrGe microcrystals.

\section*{Discussion}
\label{sec:discuss}

In conclusion, we have investigated and compared the superconducting properties of Pd, Pt, Rh, and Ir germanide thin films formed by solid-phase epitaxy, with detailed characterization of IrGe and IrSiGe films. The optimal annealing temperature increases from Pd to Ir but remains within the thermal budget of strained Ge/SiGe heterostructures. We confirmed that the mobility of the two-dimensional hole gas in strained Ge quantum wells is not degraded by the highest annealing temperatures used in this study. High-resolution STEM and EDX analyses of IrGe films reveal that Ir reacts with Ge to form a self-limiting, polycrystalline IrGe layer with a sharp IrGe/Ge interface. Optimally annealed IrGe films exhibit the highest critical temperature $T_\mathrm{c}\approx 3.4\,\mathrm{K}$ and upper critical magnetic field $B_\mathrm{c2}\approx 3.4$ T. In contrast, over-annealed IrGe films show a suppression of superconductivity, with both $T_\mathrm{c}$ and $B_\mathrm{c2}$ dropping to near zero, which we attribute to the formation of the semiconducting Ir$_4$Ge$_5$ phase. IrSiGe films exhibit slightly lower superconducting parameters, with $T_\mathrm{c}\approx 2.6\,\mathrm{K}$ and $B_\mathrm{c2}\approx 2.8\,\mathrm{T}$, but demonstrate greater thermal stability. This enhanced robustness is likely due to the formation of the more stable metallic IrSi phase. Our results establish that Ir(Si)Ge thin films formed via solid-phase epitaxy provide a viable route for fabricating high-$T_\mathrm{c}$, high-$B_\mathrm{c}$ superconducting contacts to Ge-based Josephson field-effect transistors (Ge JoFETs). These materials offer a promising platform for integrating high-performance superconducting electronics with strained Ge quantum wells in a process fully compatible with mature CMOS technology. 

\section*{Supplementary Material}

The \hyperref[supp]{supplementary material} contains detailed information on sample fabrication and characterization methods; tables that summarize the dependence of films properties on annealing conditions; and additional figures of STEM and EDX analysis of optimally annealed Ir/Ge films. 

\begin{acknowledgments}
H.L.~and L.P.R.~were supported by NSF Award \#2005092. M.M.~acknowledges the financial support by EPSRC project EP/X039757. Ir deposition was supported by the Center for Integrated Nanotechnologies through the User Proposal \#2023BU0040. This work was performed, in part, at the Center for Integrated Nanotechnologies, an Office of Science User Facility operated for the U.S.~Department of Energy (DOE) Office of Science. Sandia National Laboratories is a multimission laboratory managed and operated by National Technology \& Engineering Solutions of Sandia, LLC, a wholly owned subsidiary of Honeywell International, Inc., for the U.S.~DOE’s National Nuclear Security Administration under contract DE-NA-0003525. The views expressed in the article do not necessarily represent the views of the U.S.~DOE or the United States Government. 
\end{acknowledgments}

\section*{Author Declarations}

\subsection*{Conflict of Interest}

The authors have no conflicts to disclose. 

\subsection*{Author Contributions}

\textbf{Hao Li}: Conceptualization (equal); Data Curation (lead); Formal Analysis (lead); Investigation (lead); Validation (lead); Visualization (lead); Writing - Original Draft Preparation (lead); Writing - Review \& Editing (equal). \textbf{Zhongxia Shang}: Investigation (equal); Writing - Review \& Editing (supporting). \textbf{Michael P. Lilly}: Resources (equal); Writing - Review \& Editing (supporting). \textbf{Maksym Myronov}: Resources (equal); Writing - Review \& Editing (supporting). \textbf{Leonid P. Rokhinson}: Conceptualization (lead); Formal Analysis (equal); Funding Acquisition (lead); Methodology (lead); Project Administration (lead); Supervision (lead); Visualization (equal); Writing - Original Draft Preparation (equal); Writing - Review \& Editing (lead).

\section*{Data Availability}

The data that supports the findings of this study are available within the article and its \hyperref[supp]{supplementary material}. 

\clearpage
\newpage
\onecolumngrid

%-------------------------------------------------------------------------------------
% Supplementary material
%-------------------------------------------------------------------------------------

\renewcommand{\thefigure}{S\arabic{figure}}
\renewcommand{\theequation}{S\arabic{equation}}
\renewcommand{\thetable}{S\arabic{table}}
\renewcommand{\thepage}{sup-\arabic{page}}
\setcounter{page}{1}
\setcounter{equation}{0}
\setcounter{figure}{0}
\setcounter{table}{0}
\setcounter{section}{0}

\begin{center}
\textbf{\Large Characterization of superconducting germanide and germanosilicide films of Pd, Pt, Rh and Ir formed by solid-phase epitaxy}\\
\vspace{0.5cm}
\textbf{\large Supplementary Material} \\
\label{supp}
{\it Hao Li, Zhongxia Shang, Maksym Myronov and Leonid P. Rokhinson}
\end{center}

\tableofcontents

\clearpage

The supplementary material contains detailed information on sample fabrication and characterization methods; tables that summarize the dependence of films properties on annealing conditions; and additional figures of STEM and EDX analysis of optimally annealed Ir/Ge films. 

\section{Methods}
\label{supp:methods}

\SG\ and \textit{p}-Ge wafers used in this study were grown on \textit{p}-doped Si(001) substrates by reduced-pressure chemical vapor deposition (RP-CVD). Undoped Ge(001) wafers were purchased from MTI Corporation. All wafers were degreased using toluene, acetone, and isopropanol with ultrasonication. Prior to loading into the deposition system, the native surface oxide was removed by sequential dips in HCl:H$_2$O (1:6) for 15 seconds and HF:H$_2$O (1:10) for 10 seconds, followed by a 10-second rinse in deionized (DI) water.

Iridium (Ir) was deposited by DC sputtering in a KJL PVD-75 system at Sandia National Laboratories under a 3 mTorr Ar atmosphere, using 100 W DC power, substrate rotation at 20 rpm, and a 75$^\circ$ tilt angle, at a deposition rate of approximately 1 \AA/s. Palladium (Pd), platinum (Pt), and rhodium (Rh) films were deposited at normal incidence by electron-beam evaporation in an AJA ATC 1800-HY hybrid deposition system at Purdue University, with a base pressure below $10^{-8}$ Torr and a deposition rate of 1–1.5 \AA/s. In both systems, substrates were maintained at room temperature during deposition.

Following deposition, samples were annealed in an Ar atmosphere using a home-built rapid thermal annealing system with a heating and cooling rate of $1\ ^\circ$C/s. Electrical and magnetotransport measurements were conducted on samples patterned in either Hall bar or van der Pauw geometries, using a $^3$He system or a dilution refrigerator. Conventional low-frequency (10–30 Hz) four-terminal lock-in techniques were employed.

For high-resolution scanning transmission electron microscopy (HRSTEM) studies, a $<100$ nm-thick lamella was prepared using a Thermo Scientific Helios G4 UX DualBeam focused ion beam (FIB) system. Transmission electron microscopy (TEM) imaging and energy-dispersive X-ray spectroscopy (EDX) were performed using a Thermo Scientific Themis Z system at an acceleration voltage of 300 kV. Atomic force microscopy (AFM) images were acquired using a Veeco Dimension 3100 AFM system.

\clearpage
\section{Extended tables}
\label{supp:tables}

\begin{table*}[h]
\begin{threeparttable}
\begin{center}
\caption{Properties of optimally annealed MGe and MSiGe films, where M = Pd, Pt, Rh or Ir, $d^\mathrm{M}$ is the metal thickness, $T_\mathrm{ann}$ is annealing temperature, and $t_\mathrm{ann}$ is annealing time. The superconducting transition widths for critical temperature $\Delta T_\mathrm{c}=T_\mathrm{c}^{95}-T_\mathrm{c}^5$ and for out-of-plane critical field $\Delta B_\mathrm{c2,\perp}=B_\mathrm{c2,\perp}^{95}-B_\mathrm{c2,\perp}^{5}$, where $T_\mathrm{c}^{XX}$ and $B_\mathrm{c2,\perp}^{XX}$ refers to temperatures and out-of-plane fields where resistance reaches $XX\%$ of it's normal value. $B_\mathrm{c2,\perp}(0)$ are critical fields extrapolated to $T=0$, see the main text. }
\label{tab:PdPtRhIrGeSiGeTcBc}
\begin{tabularx}{\textwidth}{c@{\hspace{0.04\textwidth}}>{\centering\arraybackslash}X>{\centering\arraybackslash}X>{\centering\arraybackslash}X>{\centering\arraybackslash}X@{\hspace{0.08\textwidth}}>{\centering\arraybackslash}X>{\centering\arraybackslash}X>{\centering\arraybackslash}X}
\hline\hline
Substrate & \multicolumn{4}{c@{\hspace{0.08\textwidth}}}{Ge (001)} & \multicolumn{3}{c}{\SG}\\
M & Pd & Pt & Rh & Ir & Pd & Pt & Ir\\
\hline
$d^\mathrm{M}$ (nm) & 10 & 10 & 10 & 10 & 20 & 20 & 10\\
$T_\mathrm{ann}$ ($^\circ\mathrm{C}$) & 400 & 400 & 500 & 450 & 400 & 350 & 475\\
$t_\mathrm{ann}$ (min) & 10 & 5 & 5 & 30 & 10 & 5 & 20\\
\rule{0pt}{4ex}
$T_\mathrm{c}$ (K) & 0.157 & 0.500 & 1.68 & 3.44 & 0.299 & 0.585 & 2.56\\
$\Delta T_\mathrm{c}$ (K) & 0.03 & 0.19 & 0.28 & 0.38 & 0.01 & 0.02 & 0.31\\
\rule{0pt}{4ex}
$B_\mathrm{c2,\perp}(0)$ (T) &  &  & 1.46 & 3.43 &  &  & 2.84\\
$\Delta B_\mathrm{c2,\perp}$ (T)\tnote{a)} &  &  & 0.30 & 1.38 &  &  & 0.73\\
\hline\hline
\end{tabularx}
\begin{tablenotes}
\item[a)] Measured at $T=0.25$ K. 
\end{tablenotes}
\end{center}
\end{threeparttable}
\end{table*}

\clearpage

\begin{table*}[t]
\begin{threeparttable}
\begin{center}
\caption{Parameters of IrGe (top) and IrSiGe (bottom) films formed using different annealing conditions. Most parameters are defined in \hyperref[tab:PdPtRhIrGeSiGeTcBc]{Table.~S1}. Sheet resistivity $\rho_\square$ is measured in van der Pauw geometry. Resistivity $\rho$ and electron density $n$ are calculated assuming that all Ir reacts with Ge to form IrGe with the thickness $2.37\ d^\mathrm{Ir}$. }
\label{tab:IrGeIrSiGeTcBc}
\begin{tabularx}{\textwidth}{>{\centering\arraybackslash}X>{\centering\arraybackslash}X>{\centering\arraybackslash}X>{\centering\arraybackslash}X>{\centering\arraybackslash}X>{\centering\arraybackslash}X>{\centering\arraybackslash}X>{\centering\arraybackslash}X>{\centering\arraybackslash}Xc@{\hspace{0.03\textwidth}}>{\centering\arraybackslash}Xcc}
\hline\hline
$d^\mathrm{Ir}$ & $T_\mathrm{ann}$ & $t_\mathrm{ann}$ & $T_\mathrm{c}$ & $\Delta T_\mathrm{c}$ & $B_{\mathrm{c2,\perp}}$ & $\Delta B_{\mathrm{c2,\perp}}$ & $\rho_\square$ & $\rho$ & $n$ \\
$(\mathrm{nm})$ & $(^\circ\mathrm{C})$ & $(\mathrm{min})$ & $(\mathrm{K})$ & $(\mathrm{K})$ & $(\mathrm{T})$ & $(\mathrm{T})$ & $(\Omega/\square)$ & $(\mu\Omega$-$\mathrm{m})$ & $\mathrm{10^{27}m^{-3}}$ \\
\hline
\bf{\textit{i}-Ge} & 400 & 2 & 0.68 & 0.09 & 0.34 & 0.16 & 136 & \textbf{--} & \textbf{--} \\
\multirow{2}{*}{5} & 425 & 2 & 1.00 & 0.08 & 0.64 & 0.17 & 134 & \textbf{--} & \textbf{--} \\
 & 450 & 2 & 2.72 & 0.14 & 1.96 & 0.33 & 117 & 1.39 & \textbf{--} \\
 & 475 & 2 & 2.92 & 0.16 & 2.33 & 0.46 & 115 & 1.36 & \textbf{--} \\
\rule{0pt}{5ex}
\multirow{5}{*}{10} & 450 & 30\tnote{a)} & 3.44 & 0.31 & 2.90 & 1.39 & 209 & 4.95 & 1.88 \\
 & 475 & 3.5 & 3.30 & 0.23 & 2.19 & 0.58 & 44.5 & 1.05 & 13.9 \\
 & 475 & 3.5 & 3.44 & 0.19 & 2.39 & 0.76 & 52.0 & 1.23 & \textbf{--} \\
 & 500 & 1 & 3.10 & 0.84 & 3.07 & 2.10 & 217 & 5.14 & \textbf{--} \\
 & 500 & 2\tnote{b)} & $<$0.25 & \textbf{--} & \textbf{--} & \textbf{--} & 430 & \textbf{--} & \textbf{--} \\
\hline
\rule{0pt}{3ex}
\bf{SiGe} & 450 & 20 & 1.99 & 0.23 & 1.57 & 0.45 & 64.9 & 1.54 & \textbf{--} \\
\multirow{1}{*}{10} & 475 & 10 & 2.50 & 0.21 & 2.30 & 0.65 & 59.5 & 1.41 & \textbf{--} \\
 & 475 & 20\tnote{a)} & 2.61 & 0.32 & 2.37 & 0.76 & 64.2 & 1.52 & \textbf{--} \\
\hline\hline
\end{tabularx}
\begin{tablenotes}
\item[a)] These conditions are used in the main text analysis. 
\item[b)] For this annealing condition there is an onset of a broad transition with $T_\mathrm{c}^{95}=1.00\,\mathrm{K}$ and $B_\mathrm{c2,\perp}^{95}=2.71\,\mathrm{T}$. 
\end{tablenotes}
\end{center}
\end{threeparttable}
\end{table*}

\clearpage
\section{Extended figures}

\begin{figure*}[h]
\centering
\includegraphics[width=0.7\textwidth]{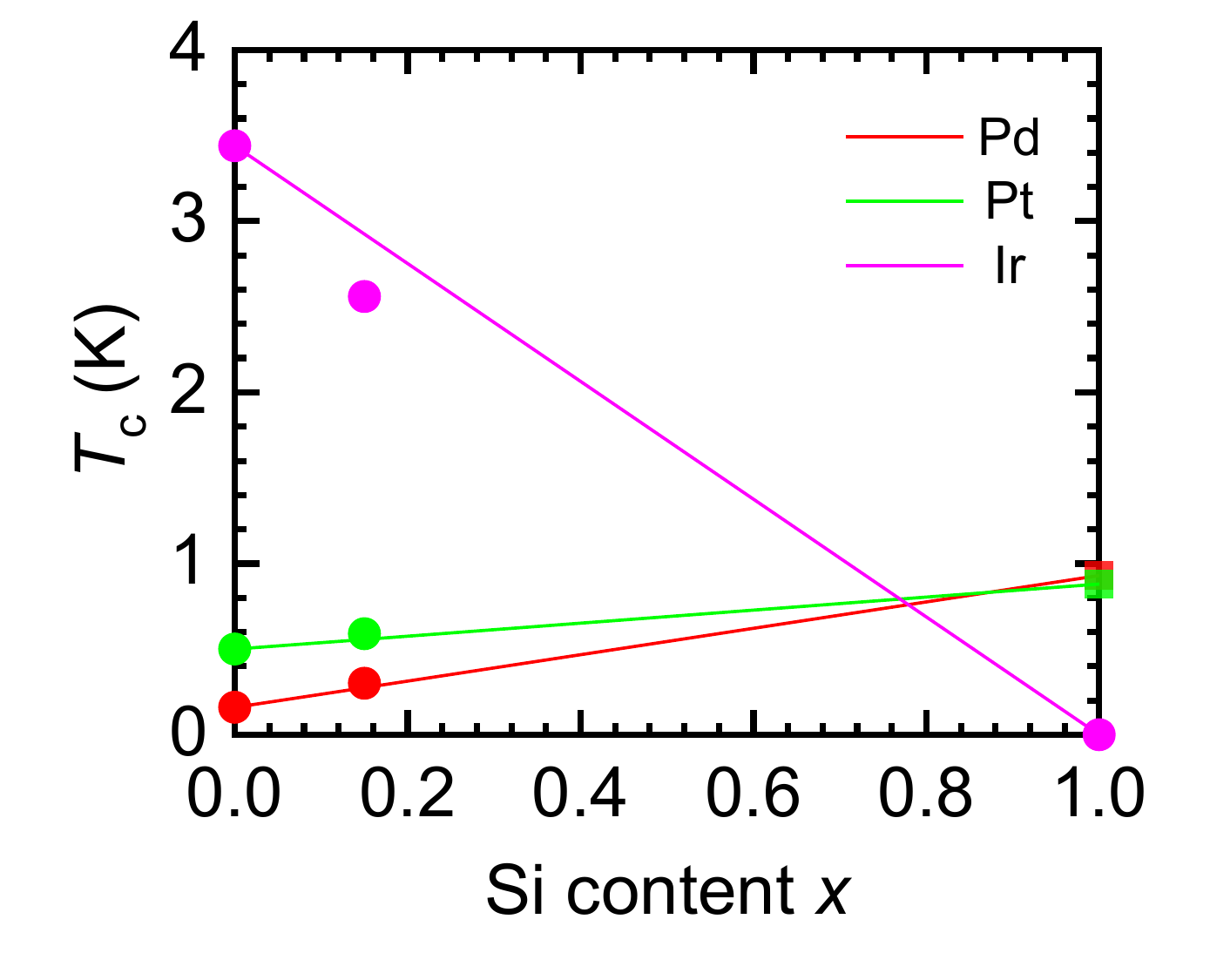}
\caption{$T_\mathrm{c}$ for pure MSi are shown by squares symbols \cite{Matthias1963,Raub1963,Roberts1976,Raub1984}. Measured $T_\mathrm{c}$ for MGe, MSi and $\mathrm{MSi_{0.15}Ge_{0.85}}$ are plotted as dots. These values are close to the values linearly extrapolated between $T_\mathrm{c}$ of MSi and MGe.}
\label{fig:PdRhIrSixGe1-xTc}
\end{figure*}

\clearpage

\begin{figure*}[t]
\centering
\includegraphics[width=0.7\textwidth]{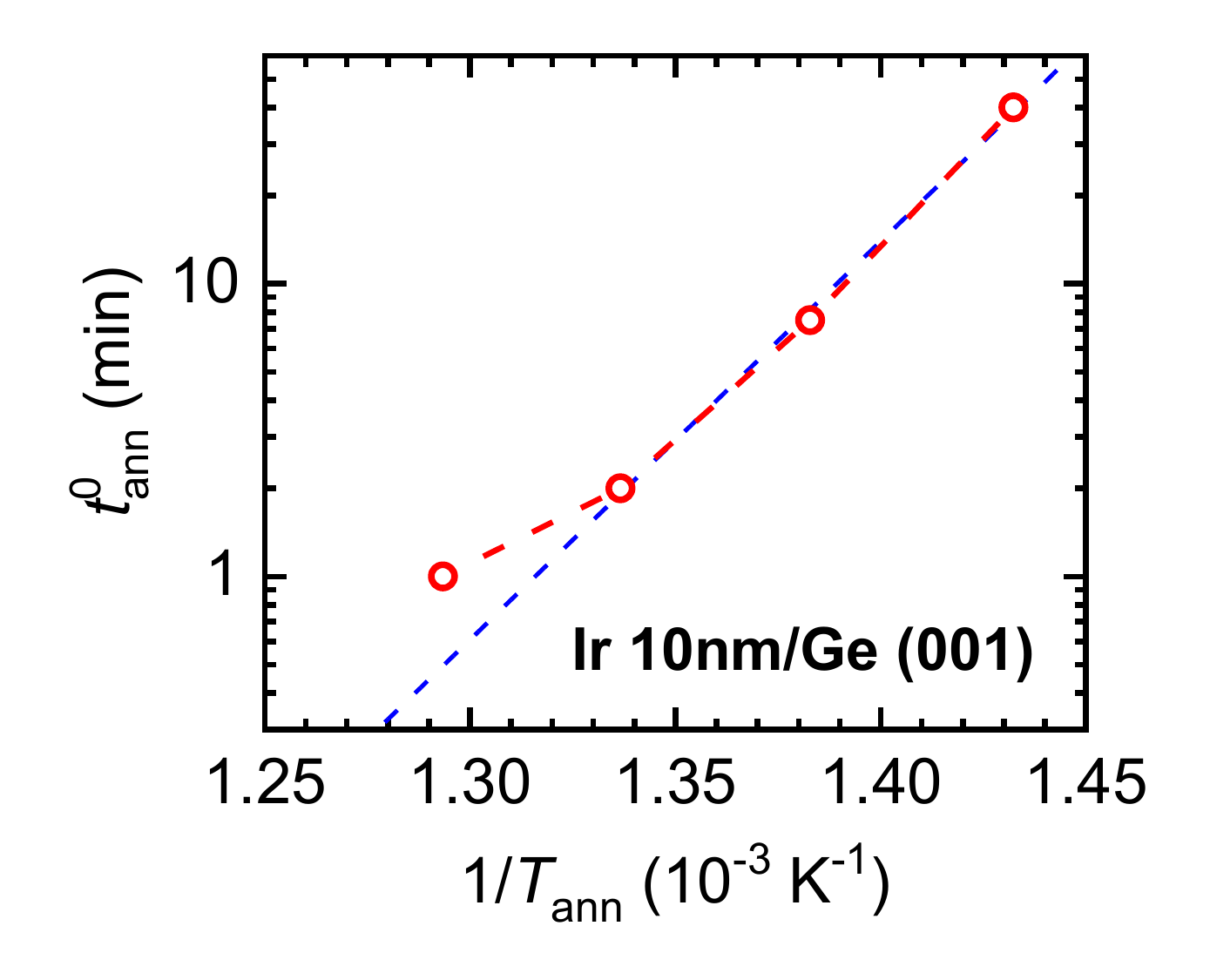}
\caption{The Arrehenius plot of $t^0_\mathrm{ann}$ vs. $1/T_\mathrm{ann}$ for Ir $10\,\mathrm{nm}$/\textit{i}-Ge films. $t^0_\mathrm{ann}$ is defined as the annealing time required for the film to have a critical temperature $T_\mathrm{c}\sim(2.8\pm0.2)\,\mathrm{K}$. The dashed line corresponds to the activation energy $2.7$ eV.}
\label{fig:IrGeArrehenius}
\end{figure*}

\clearpage

\begin{figure*}
    \centering
    \includegraphics[width=0.7\textwidth]{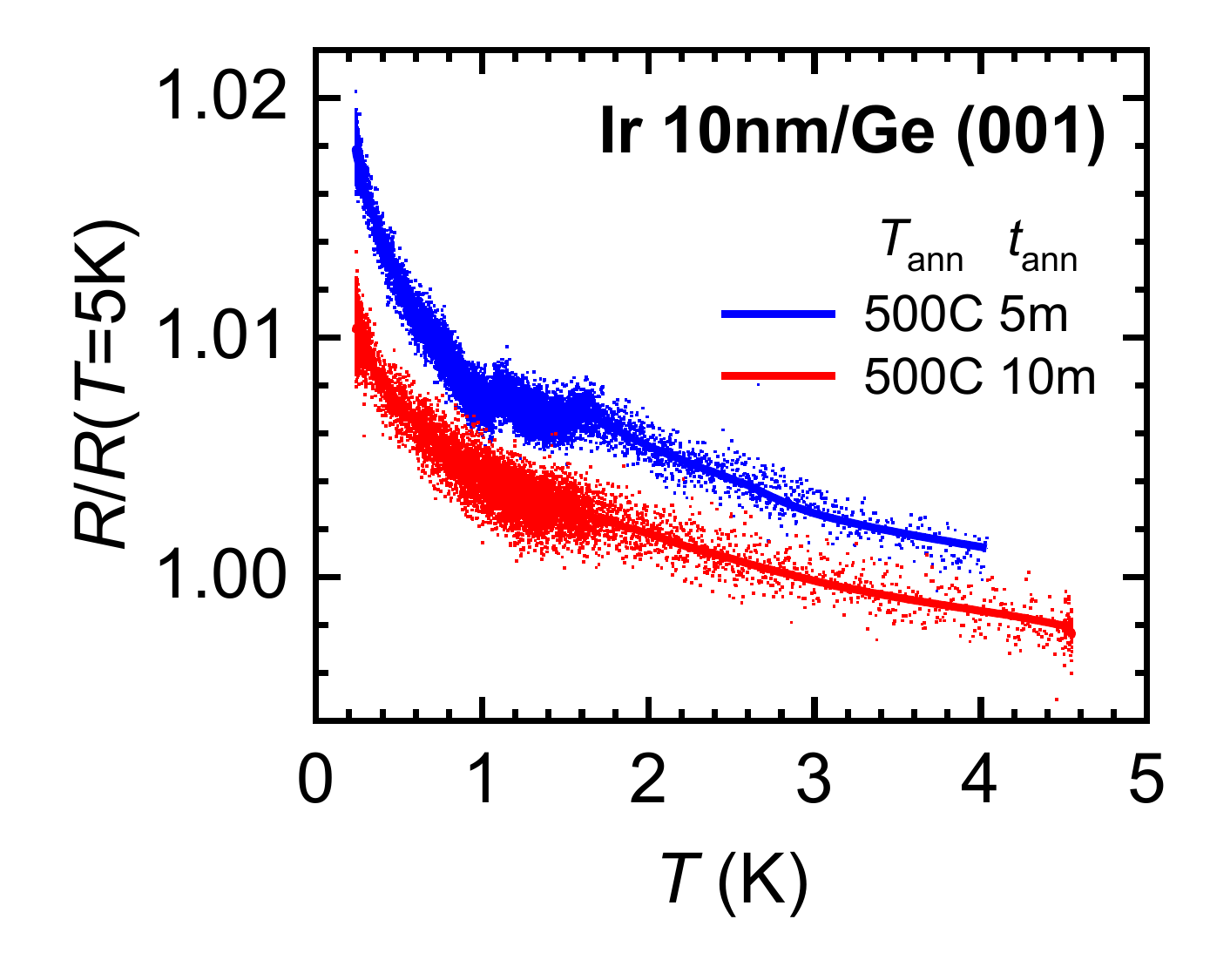}
    \caption{Temperature dependence of normalized resistance in over-annealed Ir/\textit{i}-Ge films shows a negative trend at low temperatures, indicating semiconducting properties of the film.}
    \label{fig:IrGeoverannRT}
\end{figure*}

\clearpage

\begin{figure*}[htbp]
\includegraphics[width=\textwidth]{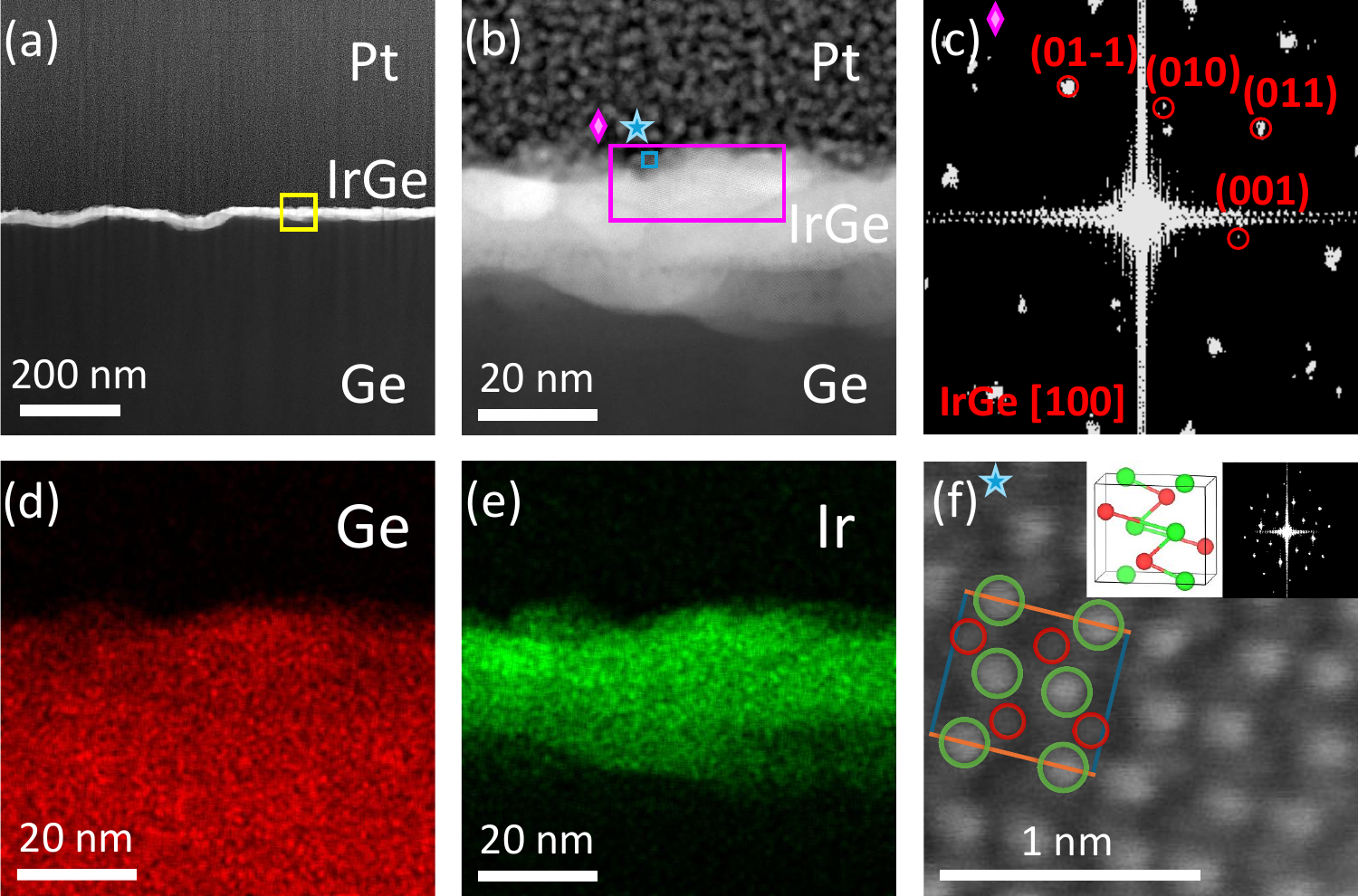}
\caption{(a) Low-magnification cross-sectional STEM image of an Ir $10\,\mathrm{nm}$/\textit{i}-Ge film annealed at $475\,^\circ\mathrm{C}$ for $3.5\,\mathrm{min}$. The sample is coated with a Pt protection layer. (b) HAADF-STEM image of a flat region of the IrGe film (yellow square in (a)), with (c) the indexed FFT (of magenta region in (b)) and (d,e) corresponding EDX maps. (f) Atomic-resolution STEM image near the top surface of the film (blue square in (b)), showing the orthorhombic phase (Pnma, space group No.\ 62) of IrGe, taken along the [100] orientation of the IrGe grain. Insets: schematic of the IrGe lattice and FFT of (f). Green circles mark Ir atoms and red circles mark Ge atoms. The $b$ and $c$ axes of the IrGe lattice are highlighted by blue and orange lines, respectively.}
\label{fig:IrGeSTEM100}
\end{figure*}

\bibliography{IrGeTcBc}

\end{document}